\documentclass[aps,prl,superscriptaddress,showpacs,floatfix,twocolumn]{revtex4-1}
\usepackage{times}
\usepackage{graphicx,graphics,color}
\usepackage{amsmath}
\usepackage{amssymb}
\usepackage{ulem} 
\usepackage{color}

\begin{document}

\title{Phase diagram of doped BaFe$_2$As$_2$ superconductor under broken $C_4$ symmetry}

\author{Yuan-Yen Tai}
\affiliation{Department of Physics, University of Houston, Houston, Texas 77004, USA}
 
\author{Jian-Xin Zhu}
\affiliation{Theoretical Division, Los Alamos National Laboratory, Los Alamos, New Mexico 87545, USA}

\author{Matthias J. Graf}
\affiliation{Theoretical Division, Los Alamos National Laboratory, Los Alamos, New Mexico 87545, USA}

\author{C. S. Ting}
\affiliation{Department of Physics, University of Houston, Houston, Texas 77004, USA}

\date{\today}

\begin{abstract}
We develop a minimal multiorbital tight-binding model with realistic hopping parameters. The  model breaks the symmetry of the tetragonal point group by lowering it from $C_4$ to $D_{2d}$, which accurately describes the Fermi surface evolution of the electron-doped BaFe$_{2-x}$Co$_x$As$_2$ and hole-doped Ba$_{1-y}$K$_y$Fe$_2$As$_2$  compounds. 
An investigation of the phase diagram with a mean-field $t$-$U$-$V$ Bogoliubov-de Gennes Hamiltonian results in agreement with the experimentally observed electron- and hole-doped phase diagram with only one set of $t$, $U$ and $V$ parameters. Additionally, the self-consistently calculated superconducting order parameter exhibits $s^\pm$-wave pairing symmetry with a small $d$-wave pairing  admixture in the entire doping range,
which is the subtle result of the weakly broken symmetry and competing interactions in the multiorbital mean-field Hamiltonian.
\end{abstract}
\pacs{78.70.Dm, 71.10.Fd, 71.10.-w, 71.15.Qe}

\maketitle

\paragraph{Introduction.$-$} The discovery of the iron-based superconductors attracted much experimental and theoretical attention
leading to many microscopic model studies \cite{DZhang,TZhou,JXZhu2011,WeiLi:2012,WeiLi:2012a,Das2011,JiangpingHu2012,NingningHao2012,Xu2008,Kuroki2008}.
Of the various effective microscopic models, the one proposed by Zhang \cite{DZhang} accounted for the effects of the upper and lower anion atoms (above and below the Fe plane).
Zhang's minimal model successfully described the behavior of the collinear antiferromagnetism (C-AFM) \cite{PRichard2012,Cruz2008,YChen2008} and the competition with superconducting (SC) order in the electron-doped ($e$-doped) part of the phase diagram when substituting Fe with Co atoms. 
However, the model failed to give account of the hole-doped ($h$-doped) part of the phase diagram when doping on the Ba site. 
Very recently,  Hu and co-workers \cite{JiangpingHu2012} proposed a Hamiltonian with $S_4$ symmetry to clarify the local symmetry breaking of the underlying electronic structure through orbital ordering in the Fe-122 family. Both models by Zhang~\cite{DZhang} and Hu~\cite{JiangpingHu2012} share a very important ingredient, that is, the breaking of the $C_4$ symmetry. So far both models have been applied only to the $e$-doped  region of the phase 
diagram \cite{TZhou,NingningHao2012}.
Although both  models  give a qualitative picture of competing order in the system,  caution must be taken when the evolution of the low-energy quasiparticle states and Fermi surface topology are considered. These can impose even stricter constraints on a given model:   
(i) $e$-doping: The disappearance of the hole pockets at the $\Gamma$ point of the Brillouin zone (BZ) is the key feature of $e$-doped compounds BaFe$_{2-x}$Co$_x$As$_2$ \cite{YSekiba,Pratt2009,Lester2009,XWang2009,Laplace2009}. 
(ii) $h$-doping: The nested large electron pockets around the $M$ point of the BZ evolve into a set of four small clover-like hole pockets for the end member KFe$_2$As$_2$, which is identified as a Dirac cone \cite{TSato,Rotter2008,HChen2009}. A very similar feature is also found in several density functional theory (DFT) calculations \cite{JXZhu2011,WeiLi:2012a,Mazin2008,Singh2008,Xu2008}.
Note that this is an intrinsic feature of the bare band structure and is not related to emergent spin-density wave order  \cite{PRichard2012,TZhou02}.

\begin{figure}  
\includegraphics[scale=0.25,angle=0]{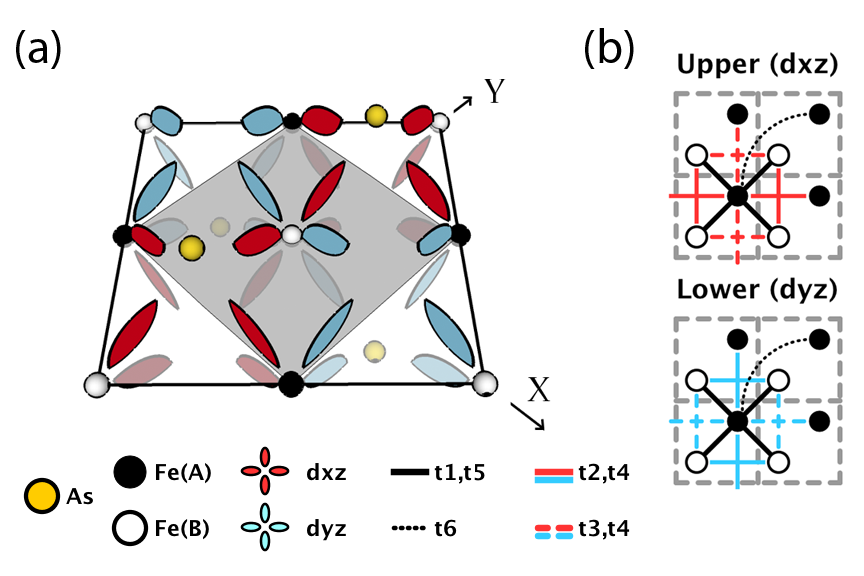}
\caption
{(color online)
(a) Schematic picture of the Fe $d_{xz}$ and $d_{yz}$  orbital overlap through the As atoms.
In panel (a), the red (blue) lobes represent the $d_{xz}$ ($d_{yz}$) orbitals. The white (black) spheres represent the Fe atoms on the A  (B) sublattice and the yellow spheres represent  the upper (lower) As atoms.
This configuration shows a 2$\times$2 (4-Fe) unit cell within the solid thick lines. The shaded region is the 2-Fe unit cell.
The 2-Fe unit cell is used in panel (b) to construct the tight-binding model, where hopping through the As atoms is ignored.
In panel (b), the solid black line stands for nearest-neighbor (1NN) intra (t$_1$) and inter (t$_5$) hopping parameters, the solid red (blue) line stands for 2NN hopping t$_2$  (t$_4$), the dashed red (blue) line is for second-nearest-neighbor (2NN) hopping t$_3$  (t$_4$), and the dotted black line stands for the third-nearest-neighbor (3NN) hopping t$_6$.
}\label{fig:model}
\end{figure}

Aided by experiments and DFT calculations,  we develop a minimal tight-binding model  with improved normal-state band structure parametrization to account for conditions (i) and (ii) .
The crystal structure of BaFe$_2$As$_2$ (ThCr$_2$Si$_2$ type structure) and
its electronic structure are now well understood and the relevant orbitals for atomic bonding have been identified \cite{CZheng1988}.
Within a quantum-chemical framework of bonding one realizes that the upper (lower) anion atom gives rise to different overlap between the Fe $3d_{xz}$  ($3d_{yz}$) orbital with the $4p$ orbitals of the As atom.
A schematic picture of the overlap between these orbitals  is shown in Fig.~\ref{fig:model}(a). The  challenge in the study of superconductivity in the Fe-based 122 family  is how to simultaneously satisfy the observations (i) and (ii).

In this Letter, we show that in order to model the phase diagram, one needs to introduce three key conditions critical for the second-nearest neighbor (2NN) hopping terms.
{\bf I:\ } On the same orbital, we break $C_4$ symmetry between the A  and B sublattices of the Fe atoms ($t_2 \neq t_3$), see Fig.~\ref{fig:model}(b).
 {\bf II:\ } In the same sublattice, we break the degeneracy between the $d_{xz}$ and $d_{yz}$ orbitals by adding a {\it twist}.
{\bf III:\ } We include the 2NN inter-orbital hopping term $t_4$.
The combined effects of conditions {\bf I} and {\bf II} reduce the symmetry to $D_{2d}$. For a detailed explanation see the Supplemental Material (SM) \cite{SM}.

\begin{figure}
\includegraphics[scale=0.55,angle=0]{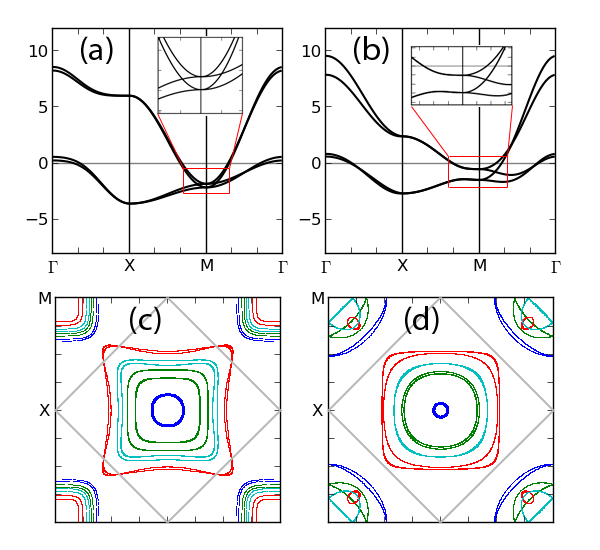}
\caption
{(color online)
The electronic band structure of model Hamiltonians H$^t_1$ (a) and H$^t_2$ (b) along high-symmetry directions in the BZ. The insets show the significant difference between H$^t_1$ and H$^t_2$ near the $M$ point. The horizontal line in (a) and (b) indicates the Fermi level $E_f$ for half filling ($n=2.0$). 
Panel (c) shows the Fermi surfaces for model H$^t_1$ for $n=2.23$ (blue, $\mu$=-0.76), $n=2.0$ (green, $\mu$=-1.244), $n=1.8$ (cyan, $\mu$=-1.65), $n=1.5$ (red, $\mu$=-2.25).
Panel (d) shows the FSs of model H$^t_2$ for different doping levels $n=2.38$ (blue, $\mu$=-0.055), $n=2.0$ (green, $\mu$=-0.89), $n=1.8$ (cyan, $\mu$=-1.25), $n=1.5$ (red, $\mu$=-1.745).
}\label{fig:band}
\end{figure}

\paragraph{Theory.$-$} The original set of four hopping parameters, $t_{1-4}$, in Zhang's model accounted only for conditions {\bf I} and {\bf III}.
In our new model, we include the symmetry condition {\bf II} and thus extend the set to six hopping parameters $t_{1-6}$.
For the calculations we choose the 2-Fe unit cell, as described in Fig.~\ref{fig:model}(a), to construct the basis function 
$\psi$ = $(c_{A1}, c_{A2}, c_{B1}, c_{B2})^T$ 
and its Fourier transform 
$c_{iM\alpha}=\frac{1}{\sqrt{N}}\,\sum_{{\bf k}} c_{M\alpha}({\bf k})\times\exp\big( i {\bf R}_{iM}\cdot {\bf k} \big)$ 
for the normal state $k$-space.
The  kinetic energy is then given by the minimal hopping Hamiltonian 
$H^t = \sum_{{\bf k}}\psi^\dagger({{\bf k}})\, W_{{\bf k}}\, \psi({{\bf k}})$ 
in the 2-Fe unit cell of the 1$^{st}$ BZ, $-\pi< k_x(k_y) <\pi$, where M=(A,B) and ${\bf R}_{iM}$ is the position of the Fe atoms in the A (B) sublattice, $\alpha$=(1,2) stands for $d_{xz}$ / $d_{yz}$ orbitals, and
\begin{equation}
\begin{aligned}
\label{wk}
W_{{\bf k}} =& 
\left(
\begin{array}{cccc}
 \xi_{A1}-\mu& \xi_{12}& \xi_{t}&  \xi_{c} \\
 \xi_{12}& \xi_{A2}-\mu& \xi_{c}&  \xi_{t} \\
 \xi_{t}&  \xi_{c}&  \xi_{B1}-\mu& \xi_{12} \\
 \xi_{c}&  \xi_{t}&  \xi_{12}& \xi_{B2}-\mu\\
\end{array}
\right) .
\end{aligned}
\end{equation}

First, we compare Zhang's minimal model H$^t_1$, which satisfies only conditions {\bf I} and {\bf III}, with our extended  model H$^t_2$,
which satisfies conditions {\bf I} through {\bf III} \cite{footnote01}:
\begin{equation}
\begin{aligned}
	{\bf H^t_1}:\;\;\;  &\xi_{A1}=\xi_{A2}=\xi^{H}, \;\;\;\;\xi_{B1}=\xi_{B2}=\xi^{V},\\
	  			\mbox{with\ }& t_{1-6}=(-1, -0.4, 2, -0.04, 0, 0).\\ 
	{\bf H^t_2}:\;\;\;  &\xi_{A1}=\xi_{B2}=\xi^{H}, \;\;\;\;\xi_{A2}=\xi_{B1}=\xi^{V},\\
				\mbox{with\ }& t_{1-6}=(-1, 0.08, 1.35,-0.12, 0.09, 0.25).
\end{aligned}
\end{equation}
Here, we defined the dispersion functions:
$\xi^{H}  = 2 t_2 \cos(k_x)+2 t_3\cos(k_y)+4 t_6\cos(k_x)\cos(k_y)$,
$\xi^{V}  = 2 t_3 \cos(k_x)+2 t_2\cos(k_y)+4 t_6\cos(k_x)\cos(k_y)$,
$\xi_{12} = 2 t_4 \cos(k_x) +2 t_4\cos(k_y)$,
$\xi_{t } = 4 t_1 \cos(k_x/2)\cos(k_y/2)$,
$\xi_{c } = 4 t_5 \cos(k_x/2)\cos(k_y/2)$.
Fig.~\ref{fig:model}(b) shows the schematic picture of H$^t_2$. The key difference between both models is  the orientation of each orbital, which  is identical in the A and B sublattices in Zhang's model, while it has a $90^\circ$ relative rotation ({\it twist}) in the new model.
The parameters for $H^t_1$ are taken from Ref.~\onlinecite{TZhou}.
In the case of $H^t_2$ we determine the hopping parameters by comparing the calculated FS topologies to the angle-resolved photoemission (ARPES) experiments \cite{TSato} for  electron fill factors $n=2.0$ (half filling), $n=1.8$ (optimal $h$-doping) and $n=1.5$ (KFe$_2$As$_2$).
In Fig.~\ref{fig:band}(a) and (b) we plot the band dispersion for H$^t_1$ and H$^t_2$, respectively.
Note the significant difference between the electronic dispersions of these Hamiltonians, which is seen in the insets of Fig.~\ref{fig:band}(a) and (b), where we enlarge the relevant area around the $M$ point in the BZ.  In the case of (b) a linear dispersion (Dirac cone) can be found for model H$^t_2$, while none exists for model H$^t_1$ in (a).
As will become clearer, this key difference between Zhang's H$^t_1$ model and the new H$^t_2$ model is a direct consequence of condition {\bf II}.

In the next step, we construct the minimal mean-field Hamiltonian in the weak-coupling regime.
Since the C-AFM will enlarge the real-space unit cell, we need to choose the 4-Fe unit cell configuration as shown in Fig.~\ref{fig:model}(a), with the real-space Hamiltonian, $H=H^t+H^{\Delta}+H^{int}$.
Where $H^t$ is the hopping term, $H^\Delta$ is the 2NN intra-orbital pairing interaction and $H^{int}$ is the interaction term, 
which includes the Coulomb interaction $U$ and Hund's coupling $J_H$.
For details of the construction of the Hamiltonian see the SM \cite{SM}.

\paragraph{Model results.$-$} We calculate the phase diagram for both Hamiltonians $H_i$=$H^t_i+H^{int}+H^{\Delta}$ with $i=1,2$ to investigate the stability of the reported C-AFM, SC, and coexisting C-AFM/SC phases as a function of doping parameter.
The evolution with doping is described by the same set of hopping, interaction, and pairing parameters for each model across the entire doping range.
For model $H_1$ we use $(U,J_H,V)=(3.4, 1.3, 1.2)$, while for $H_2$ the set of parameters
$(U,J_H,V)=(3.2, 0.6, 1.05)$ is used \cite{footnote02}.

\begin{figure}  
 \includegraphics[scale=0.5,angle=0]{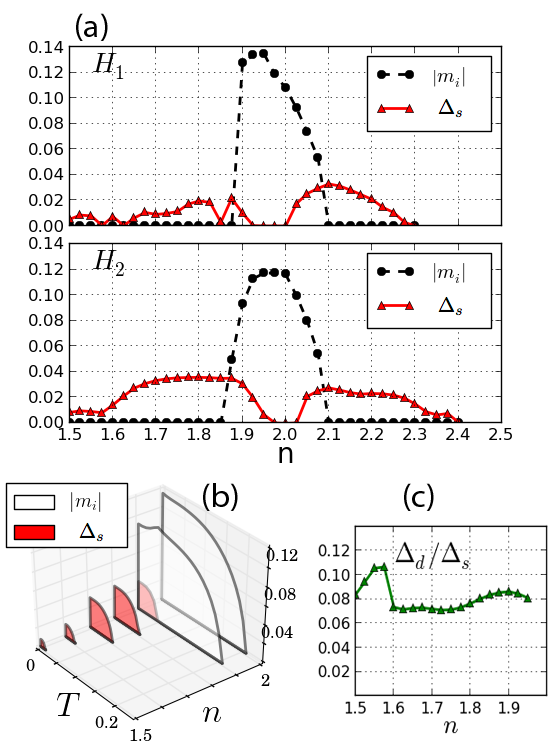}
  \caption
   {
   (color online) The phase diagrams of the C-AFM ($m_i$) and SC ($\Delta_s$) order parameters of models $H_1$ (top) and $H_2$ (bottom) at zero temperature ($T=10^{-4}$) are shown in panel (a).
   Panel (b): Temperature dependence of the calculated order parameters of model $H_2$. The suppression of $m_i$ is visible in the coexistence region with $\Delta_s$ at filling $n=1.9$.
   Panel (c): The ratio of the SC order parameters $\Delta_d/\Delta_s$ for  2NN\ pairing is shown for $H_2$. The $d$-wave admixture is of order 8\% over the entire $h$-doping regime ($n=1.5 - 2.0$). }
   \label{phase}
\end{figure}


Figure~\ref{phase}(a) shows the C-AFM and SC order parameters at nearly zero temperature ($T=10^{-4}$)
for models $H_1$ (top) and $H_2$ (bottom). We confirm that the phase diagram of $e$-doped compounds is equally well described by both Hamiltonians in agreement with experiments \cite{Pratt2009,Lester2009,XWang2009,Laplace2009,Avci2012}.
However, the situation is markedly different on the $h$-doped side. Here only  $H_2$  is capable of describing experiments \cite{Rotter2008,HChen2009}
by correctly accounting for the $e$-$h$ asymmetry and the existence of a strong SC phase at low electron filling ($1.6<n<1.85$).
In Fig.~\ref{phase}(b) the finite temperature self-consistent calculations of the C-AFM and SC order parameters are shown for  $H_2$  at $h$-doping values $n=1.5, 1.6, 1.7, 1.8, 1.9$ and $2.0$ (half filling).
We also find that in the  coexistence C-AFM/SC phase both orders compete for phase space. The competition leads to a marked suppression of $m_i$, when $\Delta_s$ nucleates at a lower temperature, see Fig.~\ref{phase}(b) for fill factor $n=1.9$. 
%
From the self-consistent calculations of the phase diagram, we extract the maximum gaps: 
C-AFM gap $m_i/T_N \approx 0.116/0.211 = 0.549$ at $n=2.0$, and the SC gaps for $e$-doping
$\Delta^{exp}_s/T_c \approx 0.11/0.049=2.23$  at $n=2.1$, and $h$-doping
$\Delta^{exp}_s/T_c \approx 0.14/0.066=2.16$  at $n=1.8$ and
$\Delta^{exp}_s/T_c \approx 0.033/0.015=2.21$ at $n=1.5$.
Here we introduced $\Delta^{exp}_s \equiv 4\,\Delta_s$ as the experimentally determined tunneling gap.
The SC gap ratios are in reasonable agreement with experimental reports for various doping values,  ranging from values associated with weak- to strong-coupling pairing, $1.5 \alt \Delta^{exp} / T_c \alt 3.8$ \cite{Teague2011,XiaohangZhang2010,Yin2009,Massee2009}.
Fig.~\ref{phase}(c) shows the ratio between $\Delta_d$ and $\Delta_s$, which is roughly 8$\%$ over the entire $h$-doping region.
Although, the $s^\pm$ gap symmetry is the most widely accepted pairing symmetry for Fe-122 based SCs, we always find a small admixture of $d$-wave symmetry in our self-consistent mean-field calculations.
A weak admixture of the $d$-wave channel is not unexpected, since $H_2$ breaks the $C_4$ symmetry of the point group.  For a lattice model with $D_{2d}$ symmetry both $s$- and $d$-wave belong to the same representation and are allowed to mix. This admixture is also found in Zhang's model, $H_1$, however, there the $C_4$ symmetry breaking is only due to condition {\bf I}, while for model $H_2$ it is related to condition {\bf I} and the 2NN\ hopping terms of condition {\bf II}.


The pairing symmetry can be discussed more systematically by how it affects the spectral function observed in ARPES experiments.  Since we are interested in the entire doping range of the phase diagram, we focus only on the  $H_2$ model.
The discussion can be further simplified by neglecting the coexistence region of C-AFM/SC. In that case, we can downfold the BdG Hamiltonian 
from the 4-Fe unit cell onto the 2-Fe unit cell configuration, see Fig.~\ref{fig:model}(a).
The order parameters $\Delta_s$ and $\Delta_d$ were previously defined on the lattice.
Their Fourier transforms in the BZ of the 2-Fe unit cell are
$\Delta_s({\bf k}) = 2\,\Delta_s\,\left[ \cos(k_x)+\cos(k_y) \right]$ and
$\Delta_d({\bf k}) = 2\,\Delta_d\,\left[ \cos(k_x)-\cos(k_y) \right]$.
We can further decompose them into  sublattice (A,B) and orbital (1,2) contributions:
$\Delta_{A1}({\bf k}) = \Delta_{B2}({\bf k}) = \Delta_s({\bf k})+\Delta_d({\bf k})$, 
$\Delta_{A2}({\bf k}) = \Delta_{B1}({\bf k}) = \Delta_s({\bf k})-\Delta_d({\bf k})$.

\begin{figure}  
 \includegraphics[scale=0.3,angle=0]{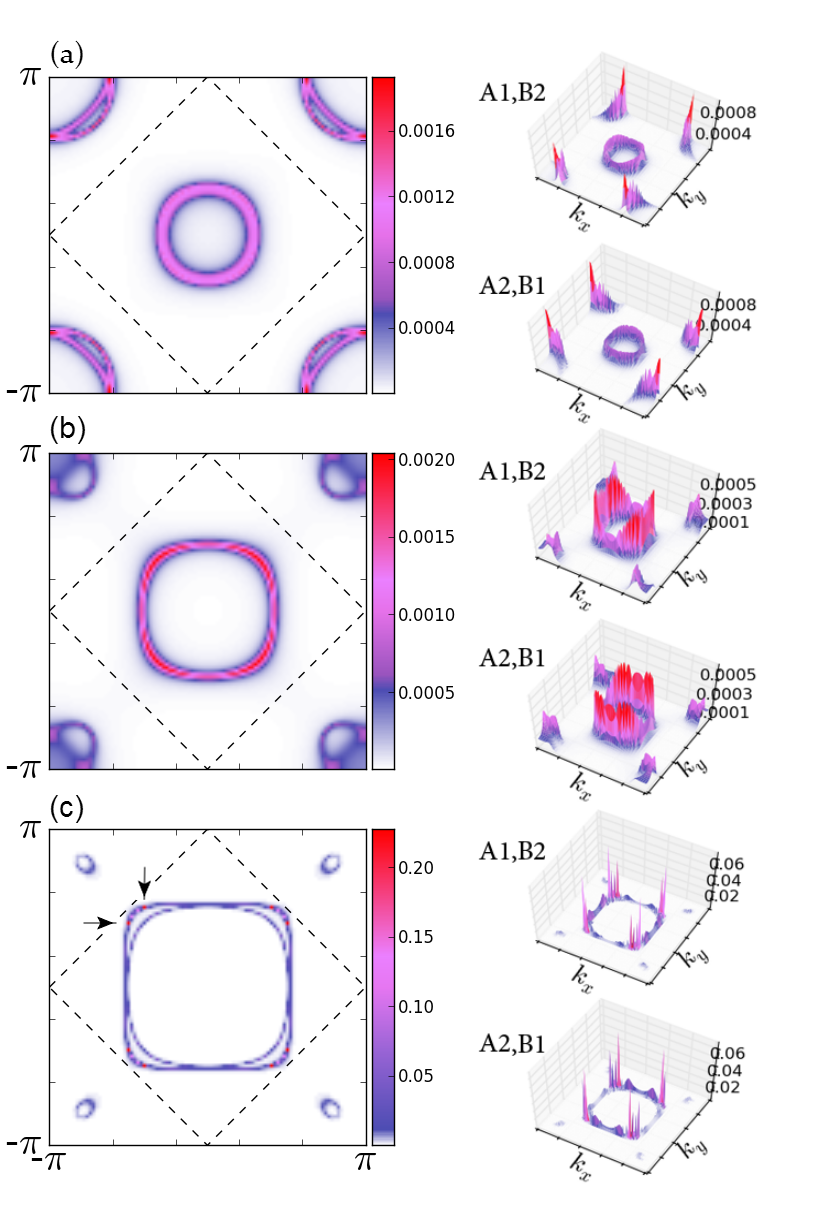}
  \caption
   {
   (color online) Spectral functions at zero temperature at the Fermi level for fill factors $n=2.1$ (a), $n=1.8$ (b), and $n=1.5$ (c). In each panel the total weight of the spectral function is shown in the left column, while the right column shows the corresponding partial spectral functions for each orbital and sublattice. The total spectrum is the sum of all four partial spectra. Panel (c): The two arrows mark the peaks of the electronic hot spots responsible for the octet gap structure. The dashed line shows the nodes of $\Delta_s({\bf k})=0$.
   }
   \label{nodal}
\end{figure}

Note these expressions are a consequence of the twofold $D_{2d}$ symmetry, thus giving rise to the s$_{x^2+y^2}$$\pm d_{x^2-y^2}$-wave gap.
The corresponding downfolded $k$-space BdG Hamiltonian attains the eigenvalues and eigenvectors for the spectral function of each orbital and sublattice,
\begin{equation}
A_{i \alpha}({\bf k}, \omega) = \sum_{n=1}^{8}  |u^{n}_{i \alpha\uparrow}(k)|^2 \delta(\omega-E^{n}_k\,)
+ |v_{i \alpha\downarrow}^{n}(k)|^2 \delta(\omega+E_{k}^{n}) ,
\end{equation}
where the total spectral function, measured in ARPES, is given by the sum $A({\bf k}, \omega) = \sum_{i,\alpha} A_{i \alpha}({\bf k}, \omega)$.
An obvious question is what is the evolution of the spectral function with doping and how is it affected by the order parameters $\Delta_s$ and $\Delta_d$ for filling factors $n=2.1, 1.8$ and $1.5$.
In Fig.~\ref{nodal} the total spectral function and its partial weights are shown. The reduced $D_{2d}$ symmetry is obvious from plots of spectral weights $A_{A1}, A_{A2}, A_{B1}$ and $A_{B2}$.
By construction of the model,  the 90$^\circ$ rotational symmetry breaking ({\it twist}) is revealed by the distinction between $A_{A1(B2)}$ and $A_{A2(B1)}$ (right panels in  Fig.~\ref{nodal}). 
Unfortunately, ARPES experiments cannot differentiate between them. However, one can compare the total spectral functions (left panels  of Fig.~\ref{nodal}) with ARPES experiments.
The spectral weights are very small for $n=2.1$ and $n=1.8$, indicating that quasiparticles are gapped.
The hole pockets around the $\Gamma$ point exhibit an isotropic gap for $n=2.1$, while an anisotropic gap with fourfold modulation is found for $n=1.8$.
On the other hand, for $n=1.5$ (corresponding to KFe$_2$As$_2$), we observe eight quasiparticle hot spots at the corners of the large hole Fermi surface centered at $\Gamma$, as well as four small pockets near the $M$ point. We believe that the hot spots  indicate  either the existence of nodal octet structure or highly anisotropic gap.

The pairing symmetry in KFe$_2$As$_2$ has been a recent topic of hot debate. The existence of only hole pockets has motivated earlier theoretical proposals of $d$-wave pairing symmetries with gap nodes \cite{RThomale:2011,SMaiti:2011}.
This would imply a change in the superconducting symmetry from $s$- to $d$-wave pairing as hole doping is increased. So far there is some experimental evidence for gap nodes \cite{HFukazawa:2009,JKDong:2010,KHashimoto:2010,JPhReid:2012}.
However,  more recent angle-resolved photoemission spectroscopy unveiled that KFe$_2$As$_2$ is a nodal $s$-wave superconductor with octet-line node structure on the large hole pockets at $k_z=\pi$ \cite{KOkazaki2012}.
Our  model calculations are consistent with an octet nodal structure on the zone-centered hole pocket at $k_z=0$.  However, the eight electronic hot spots at the corners of the large hole Fermi surface in the spectral function do not rule out a highly anisotropic gap.
Future ARPES measurements at $k_z=0$ may resolve the current disagreement about the location of  hot spots vs.\ nodal points between our results and  experiments.

\paragraph{Summary.$-$} We have shown how the $C_4$ symmetry breaking, involving the As atoms below and above the Fe layer, leads to a natural extension of Zhang's minimal model for the Fe-122 superconductors.
The new tight-binding model is in better agreement with experiments and density functional theory calculations over the entire doping range by incorporating an orbital {\it twist}.
Within the weak-coupling theory of superconductivity, the calculated phase diagram reproduces qualitatively the $e$-$h$ doping asymmetry reported in many experiments and the gap-over-$T_c$ ratios.
In addition, we always find a small $d$-wave admixture of roughly 8\% to a dominant $s$-wave SC order parameter. This admixture can give rise to quasiparticles in the spectral function resembling nodal excitations  
at electronic hot spots in the spectral function of KFe$_2$As$_2$. 
Finally, the new minimal model is computationally more efficient than similar five-band models for studying disorder effects in real space around impurities or magnetic vortices.

\acknowledgments

We thank A. V. Balatsky, T. Das, and W. Li for many helpful discussions.
This work was supported in part by the Robert A.\ Welch Foundation under Grant No.\ E-1146 (Y.-Y.T. and C.S.T.)
and through the UC Laboratory Fees Research program at LANL under the U.S.\ DOE Contract No.~DE-AC52-06NA25396 (Y.-Y.T., J.-X.Z.\ and M.J.G.).  
Y.-Y.T.\ thanks LANL for its hospitality during his visit.


%

\newpage

\section{Supplemental Material}

Here we present the construction of the mean-field Hamiltonian on a two-dimensional lattice employed in the main text.

\section{The mean-field Bogoliubov-de Gennes equation}
For the mean-field Hamiltonian, $H=H^t+H^{\Delta}+H^{int}$,
the hopping Hamiltonian $H^t$ is expressed by
\begin{equation}
H^t=\sum_{i\alpha j\alpha' \sigma} (t_{i\alpha j\alpha'} c^\dagger_{i\alpha\sigma} c_{j\alpha'\sigma}+h.c)-\mu\sum_{i\alpha\sigma}c^\dagger_{i\alpha\sigma}c_{i\alpha\sigma},
\end{equation}
where $i$ and $j$ are site indices of each Fe site, $\alpha$=(1,2) is the orbital index, $\sigma=(\uparrow,\downarrow)$ is the spin index and $\mu$ is the chemical potential.
The second term in $H$ is the mean-field superconducting (SC) pairing Hamiltonian,
\begin{equation}
H^\Delta=\sum_{ij\alpha \sigma}(\Delta_{ij\alpha} c^\dagger_{i\alpha\sigma} c_{i\alpha\bar\sigma}+h.c) .
\end{equation}
The third  term in $H$ is the on-site Coulomb interaction Hamiltonian including Hund's coupling. Following Ref.~\onlinecite{Oles2005} we write,
\begin{equation}
\begin{aligned}
H^{int}=&U\sum_{i\alpha\sigma\neq\bar\sigma}\langle n_{i\alpha\bar\sigma} \rangle n_{i\alpha\sigma}\\
			+&U'\sum_{i,\alpha\neq\alpha',\sigma\neq\bar\sigma} \langle n_{i\alpha\bar\sigma} \rangle n_{i\alpha'\sigma}\\
			+&(U'-J_H)\sum_{i,\alpha\neq\alpha',\sigma} \langle n_{i\alpha\sigma} \rangle n_{i\alpha'\sigma},
\end{aligned}
\end{equation}
where $n_{i\alpha\sigma}$ = $c^\dagger_{i\alpha\sigma}c_{i\alpha\sigma}$ and $U'$=$U-2J_H$.
After performing the Bogoliubov transformation of the electrons onto particle-hole excitations with 
$c_{i\alpha\sigma}=u^n_{i\alpha\sigma}\gamma_n+\sigma\,v^{n*}_{i\alpha\sigma}\gamma^\dagger_n$, 
the corresponding Bogoliubov-de Gennes (BdG) Hamiltonian can be written in matrix form and solved self-consistently,
\begin{equation}
\sum_{j\alpha'}
	\left(
	\begin{array}{cc}
	 H_{i\alpha j\alpha'\uparrow} 	& \Delta_{ij\alpha'} \\
	 \Delta^*_{ij\alpha'} 	& -H_{i\alpha j\alpha'\downarrow}
	\end{array}
	\right)\
	\left(
	\begin{array}{c}
	 u^n_{j\alpha'\uparrow}\\
	 v^n_{j\alpha'\downarrow}
	\end{array}
	\right)\
=
	E^n
	\left(
	\begin{array}{c}
	 u^n_{i\alpha\uparrow}\\
	 v^n_{i\alpha\downarrow}
	\end{array}
	\right) ,
\label{matrix}
\end{equation}
with particle-like ($u^n_{i\alpha\uparrow}$) and hole-like ($v^n_{i\alpha\downarrow}$) wave functions.

The self-consistent mean-field equations for the SC order parameter and spin-up and spin-down occupations are given by
\begin{equation}
	\Delta_{ij\alpha}=\frac{V_{ij}}{4}\sum_n (u^n_{i\alpha\uparrow}v^{n*}_{j\alpha\downarrow}+u^n_{j\alpha\uparrow}v^{n*}_{i\alpha\downarrow})
	\tanh\left( \frac{E_n}{2k_B T} \right),
\end{equation}
where 
\begin{equation}
\langle n_{i\alpha\uparrow} \rangle=\sum_n |u^n_{i\alpha\uparrow}|^2 f(E_n) ,
\end{equation}
and 
\begin{equation}
\langle n_{i\alpha\downarrow} \rangle=\sum_n |v^n_{i\alpha\downarrow}|^2 [1-f(E_n)].
\end{equation}
For simplicity, we  consider second-nearest-neighbor (2NN) intra-orbital pairing only with pairing potential $V_{ij}$=V.
To facilitate the discussion of physical observables and generating of the phase diagram, we define the staggered lattice magnetization and the $s$-wave and $d$-wave projections of the order parameter on the lattice:
\begin{equation}
	m_i = \frac{1}{4}\sum_\alpha ( \langle n_{i\alpha\uparrow} \rangle - \langle n_{i\alpha\downarrow} \rangle ),
\end{equation}
\begin{equation}
	\Delta_{s} = \frac{1}{8N}\sum_{i,\delta,\alpha}\Delta_{i,i+\delta,\alpha},
\end{equation}
\begin{equation}
	\Delta_{d} = \frac{1}{8N}|\sum_{i,\delta,\alpha}\,\epsilon_x\,\epsilon_y\,\Delta_{i,i+\delta,\alpha}|.
\end{equation}
The neighbors of site $i$ are reached by
$\delta=(\hat x+\hat y,\hat x-\hat y,-\hat x+\hat y,-\hat x-\hat y)$, 
with $\epsilon_x=\delta\cdot\hat x,\epsilon_y=\delta\cdot\hat y$;
$N$ is the number of Fe sites in the real-space lattice. We note that the $x$ ($y$) axis is aligned with the short Fe-Fe bond direction.

\begin{widetext}
\begin{figure*}  
\includegraphics[scale=0.30,angle=0]{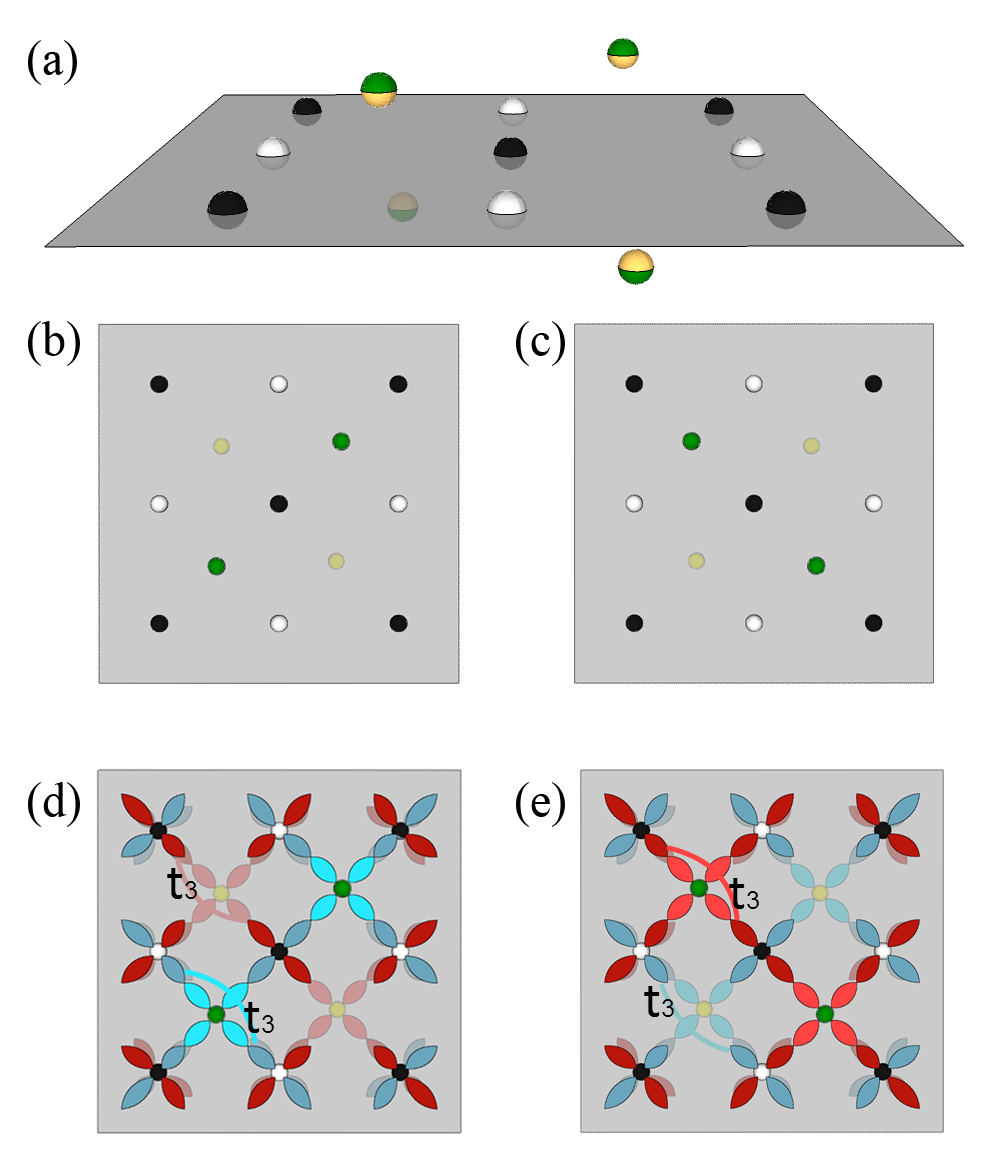}
\caption
{(color online)
(a) Schematic picture of $D_{2d}$ point group symmetry of the three-dimensional (3D) structure of the Fe-As building blocks.  The colormap is: Fe atoms  on the A (black) and B (white) sublattice;  the As atoms above (green/yellow) and below (yellow/green).
(b) Top view of panel (a);
(c) application of symmetry operation $C_4$ or $\sigma_h$ on panel (b);
(d) orbital ordering and $t_3$ hopping terms overlayed onto panel (b);
(e) the resulting $\sigma_h$ operation of panel (d).
}\label{fig:d2d}
\end{figure*}
\end{widetext}

\section{The $D_{2d}$ invariant symmetry}
We  now give the reason for our choice of the $D_{2d}$ symmetry of the point group describing the crystal structure of BaFe$_2$As$_2$, which is used
for the construction of the kinetic Hamiltonian in Eq.~(1).
To begin with, we draw the three-dimensional (3D) structure of the basic Fe-As building blocks of BaFe$_2$As$_2$ in Fig.~\ref{fig:d2d}.
The $D_{2d}$ symmetry is generated by the group elements $C_4$ and $\sigma_h$ as shown in Fig.~\ref{fig:d2d}(b,c).
Close inspection reveals the point group symmetry $D_{2d}$, because combined fourfold rotation ($C_4$) and mirror reflection ($\sigma_h$) leave the crystal structure invariant.

However, the $D_{2d}$ symmetry is obvious only for the 3D crystal structure, while the inclusion of the reflection operation $\sigma_h$ is not obvious for the two-dimensional (2D) model as shown by 
Fig.~1(a) in the main text. The question is how one can construct a $D_{2d}$ hopping Hamiltonian on a 2D lattice.
The solution is as follows: The As atoms mediate the 2NN hopping ($t_3$) through their $p$ orbitals between the $d$ orbitals of the Fe sites.
As a consequence the upper (lower) As atoms lead to effective hopping terms between the $d_{xz}$  ($d_{yz}$) orbitals, respectively.
Finally, the $\sigma_h$ operation of the upper/lower As atoms can be mapped onto the exchange of the order of the $d_{xz}$ and $d_{yz}$ orbitals, see Fig.~\ref{fig:d2d}(b). This corresponds to exchanging the upper and lower panels of Fig.~1(b) in the main text.



\begin{thebibliography}{10}
\bibitem{footnote01}%
  \BibitemOpen
  \bibinfo {note} {The Hamiltonian H$_2^t$ is invariant by exchange of orbitals
  in one sublattice if and only if one exchanges $t_1$ and $t_5$. Therefore,
  the elements in $W_k$ can be re-written: $\xi_{A1}=\xi_{B1}=\xi^H$,
  $\xi_{A2}=\xi_{B2}=\xi^V$ and $\xi_t=4\,t_5\cos(k_x/2)\,cos(k_y/2)$,
  $\xi_c=4\,t_1\cos(k_x/2)\,cos(k_y/2)$ and $\xi_{12}$ kept unchanged. In the
  main text we chose a notation to simplify the comparison to Zhang's model.}%
  \bibAnnoteFile{Stop}{footnote01}%
\end{thebibliography}

\end{document}